\documentclass[10pt]{SelfArx} 
\usepackage[english]{babel} 
\usepackage{lipsum} 
 \pdfoutput=1

\definecolor{color1}{RGB}{0,0,90} 
\definecolor{color2}{RGB}{0,20,20} 

\usepackage{hyperref} 
\hypersetup{
	hidelinks,
	colorlinks,
	breaklinks=true,
	urlcolor=color2,
	citecolor=color1,
	linkcolor=color1,
	bookmarksopen=false,
	pdftitle={Title},
	pdfauthor={Author},
}

\JournalInfo{Submitted to arXiv in August 2021} 
\Archive{Physics Education paper} 

\PaperTitle{The development of an educational software for aircraft flight mechanics calculations} 

\Authors{Claudio C. Pellegrini\textsuperscript{1}*, Mateus S. Rodrigues\textsuperscript{2}, Erika O. Moreira\textsuperscript{2}} 
\affiliation{\textsuperscript{1}\textit{Department of Thermal and Fluid Sciences, Federal University of São João del-Rei, Brazil}} 
\affiliation{\textsuperscript{2}\textit{Department of Electrical Engineering, Federal University of São João del-Rei, Brazil}} 
\affiliation{*\textbf{Corresponding author}: pelle@ufsj.edu.br} 

\newcommand{\keywordname}{Keywords} 

\Abstract{Due to its versatility and low cost, the use of unmanned aerial vehicles has been rapidly spreading in recent years, in applications ranging form military operations, to land mapping,  rescuing of lost people, aiding of natural disaster victims and many others. To properly design and operate such a vehicle, it is necessary to know its flight mechanics in the various stages of the flight. Despite the fact that the physic behind the analysis of an aircraft's flight mechanics is well known and purely based on Classical Mechanics,  the large quantity of input and output data involved favors  the use of a computational toll. This work presents the development of a toolbox called APT (Aircraft Performance Toolbox), able to make  preliminary aircraft flight mechanics calculations regarding its performance. To achieve this goal, a number of  Matlab scripts were created to perform the  calculations, and a graphical interface was created to unify them and to allow the end-user to perform the analysis in a clear and intuitive way. To illustrate the potential of the toolbox, we use APT to in the analysis of  an UAV meant to participate in the SAE Brasil AeroDesign competition of the year of 2014.}

\begin{document}

\maketitle 
\tableofcontents 
\thispagestyle{empty} 


\section{Introduction} 

The use of unmanned aerial vehicles (UAVs) has been widely disseminated over the past few years. This fact surely  stems  from its versatility: they are able to operate in different environments, weather conditions and in situations where human activity is considered  risky or even impossible \cite{saripalli}. Also, UAVs have extremely low acquisition and operating costs compared to conventional aircraft.

Among the activities in which UAVs have been intensely used, we may cite surveillance operations (military or not), the mapping of land use/land cover for geographical and meteorological purposes, the control of clandestine activities as agricultural, mineral extraction, hunting and fishing in protected areas, the control of damage caused by natural disasters and the search and rescue operations that follow, the search and rescue for people lost  in inhospitable places, as mountaineers and  sailors, the sending of supplies to isolated groups, and many others. In addiction,  UAVs may represent an important tool in the mapping of infectious diseases such as the new Covid-19. UAVs can provide spatially and temporally accurate data to help understand the links between disease transmission and environmental factors \cite{fornace}. 

Both the design and the operation of UAVs require analysis of the aircraft's flight mechanics, and in particular of its performance in various flight situations. Such analyses determine whether the aircraft is capable of fulfilling the objective for which it was designed. The idea is to be able to answer questions such as: can the UAV take off and land on the stipulated runway dimensions? Can it carry the necessary load to the prescribed distance  in the stipulated time? Can it reach the expected speed?

However, such an analysis requires input data related to the aerodynamic characteristics of the aircraft, its propulsion system, its dimensions and weight, the runway length available for takeoff and landing,  the thermodynamic properties  of the air at the operating location, etc. In addition, it inevitably ends up generating a large amount of output data.

Despite the fact that the physics used in the analysis of an aircraft's performance are well described both in the traditional and recent  literature \cite{anderson, anderson2, roskam, pellegrini_rodrigues},  the large quantity of input and output data involved favors  the use of a computational toll. For the  authors, the need for such a tool appeared during their participation in the SAE Brazil AeroDesign competition of 2014.  
The SAE Brasil AeroDesign competition, according to the organizers \cite{sae}, 
\begin{quote}
Is a challenge launched to Engineering students, whose main objective is to promote the diffusion and exchange of Aeronautical Engineering techniques and knowledge among students and future mobility engineering professionals, through practical applications and competition between teams
\end{quote}

For the competition, students must form teams composed by engineering undergraduates and a faculty advisor. The team must  design and build a cargo UAV to complete a mission that change every year. The competition aim is to disseminate knowledge in aeronautical engineering,  challenging students from all over Brazil to face each other of in an event occurring every October. 

On the year of 2009 the team Trem Ki Voa Micro was formed at the University Federal of São João del Rei, with one of the authors as its academic adviser. It made its debut at the 2010 competition and took part of every subsequent issue of the competition, until the present year. In a decade of existence the team involved more than 200 students from undergraduate courses of Mechanical, Electrical and Production Engineering,  many of whom now work in the aeronautical industry. The educational benefits and risks of this kind of competition is generally recognized by teachers and students. It is  reviewed in a number of studies (\cite{kaiser}, \cite{gadola} and \cite{schuster}for example).

With the AeroDesign competition in mind, in the year of 2014 the authors set out to obtain a computational tool to aid the team in the calculations. Their experience  showed that, in general, the computational packages developed for performance analysis must meet the following characteristics:
\begin{itemize} [noitemsep] 
\item They must be supported by mathematical models that allow a complete assessment of the aircraft's performance characteristics;
\item They must have a graphical interface that allows the creation or loading of a database containing the information of the aircraft to be analyzed, and  allowing the selection of the types of analyzes to be performed; 
\item They must generate a clear and flexible data output, allowing for the comparison of two or more aircraft; also, the resulting graphics must have layouts that can be easily adapted to the user's needs; 
\item They must be able to be implemented in a computational environment disseminated in the academic environment and be freely distributed. 
\end{itemize}

Unfortunately, a large part of aeronautical science is developed by private companies, that keep confidentiality about the tools developed by them or charge their use.  Thus, just a few computational packages developed for aircraft performance analysis are presently available and  the ones available do not meet some --- or even all --- of the requirements mentioned. Therefore, the  authors realized that the only way to assure that  the team would have access to an efficient and complete computation tool, within an affordable cost, would be to develop their own. 

This situation is not exclusive of the aircraft performance area. Thus, \cite{bloy} for example, developed an interactive multimedia package on aircraft stability and control to fulfill the gap and \cite{labfit} did the same with an statistical calculations package. 

To further illustrate the problem, the license for the ULTRA-NAV\textsuperscript{\textregistered} Performance Software\footnote{Available at http://www.ultranav.com} costs US\$ 595--795, but the package only performs take-off and landing analyses. The Atlas\textsuperscript{\textregistered} and iPreFlight\textsuperscript{\textregistered} packages\footnote{\raggedright{Available at https://www.flyapg.com/atlas.html  and  https://www.flyapg.com/ipreflight.html, respectively}},  developed by the APG (Aircraft Performance Group), perform only take-off and payload analyses respectively, and  charge a ``price on request" for their licenses. The Aircraft Performance Program\textsuperscript{\textregistered}\footnote{\raggedright{Available at http://www.darcorp.com/aircraft-performance-program-software}} developed by DARcorporation fulfills  most of the requirements mentioned above, but  its  license costs from US\$ 5,000 (intermediate version) to almost  US\$ 8,000 (full version).

The objective of the present  article is to present the design and development of the Aircraft Performance Toolbox (APT), created using the Matlab\textsuperscript{\textregistered} (student version) environment. It was designed to be a complete and free-to-use tool to help students calculate the most relevant aircraft performance parameters and  help them explore the numerous possibilities associated with different aircraft configurations. A brief account of the didactic experience with the APT in the Trem Ki Voa Micro team is also presented. 

This study is part of a larger project that includes the complete evaluation of an aircraft flight mechanics. It involves the analysis of performance presented here, but also of static stability, dynamic stability, classical feedback control, stability and control augmentation, and is currently underway.

\section{Computational tools as a teaching resource}
This is a topic that has been extensively covered in many studies in the last decades, since the popularization of personal computers started. Nevertheless, some general characteristics previously noted by some selected authors  are worth mentioning. 

Experience has shown that active visualization-based learning using interactive tolls can greatly enhance the student's understanding and retention. A direct interactive visual approach  may help to remove many of the conventional barriers that hinder the effective learning of  abstract engineering concepts \cite{wideberg}. However,  computational packages and visualisation programs are not the only possibility.  Teachers have been using non-computational tools for a long time, in a  number of creative ways, including experimental set-ups, scale models and toolkits. Extensive use has also been made in the past of video presentations by lecturers  (usually in Super 8 format)  and slideshows, using the physical photographic slides. 

Nevertheless, soon after the popularization of  personal computers, they begin to substitute the "physical'' apparatuses whenever possible, due to their flexibility and  low cost (a basic didactic wind tunnel can easily cost as much as  100,000 Euros). For example, a study by \cite{wolff} describes a numerical simulation experiments being used in the late 70's to teach basic physics  at the New York Institute of Technology. In the following decades, this tendency accelerated noticeably and at the moment, educational software is used in  almost every area of knowledge, including human sciences.  

In a  study on the use of computers as a tool in the teaching of physical sciences, \cite{mucchielle}\footnote{In Portuguese} conclude that  the pedagogical potential of computers can only be fully realized if there are enough educational programs of sufficient quality. In this respect, many papers highlight the   characteristics that an educational software must have (\cite{wideberg} and \cite{rodrigues}, for example).
The most frequent adjectives are:  interactive  programs, visualization-based learning, motivating, user-friendly interface, robustness, easy of use, and, of course, free to download and free to use.  

The didactic value of computational packages, in connection to the teaching of basic physics, is 
summarized by \cite{lopes_tort}\footnote{Translated  from Portuguese.} as: 
\begin{quote}
The development of educational software allows teachers and students to perform simulations and create animations related to  challenging problems in Newtonian kinematics and dynamics, both  in high school and in undergraduate introductory Mechanics courses. Problems that do not require the student to be familiar with advanced mathematical methods  but that allow for a wide variation of the relevant parameters, can be studied with great advantage.
\end{quote}

In conclusion, the use of  computer simulations and computer experiments in the teaching of sciences is very promising.  However, as noted by \cite{wideberg}, implementing the idea may mean  modifying not only the way of teaching but also reconsidering the educational model and philosophy.

\section{Aircraft performance}
 The performance characteristics of an aircraft requires several sets of calculations, hereafter denominated \emph{performance analyses} or {analyses} for short. Each  contains a number of computations based on Classical Mechanics, requires a considerable amount of input data and returns an equally considerable amount of output data (39 variables to be precise). 

This section briefly introduces the input variables and describes each of the analyses that compose the APT toolbox. The aeronautical terms used are traditional in the literature and may be found in aviation terms dictionaries or glossaries, as  \cite{dicionario_aviacao} for example.  

\subsection{Input variables}

The input variables of the APT can be divided into three categories: take-off variables, geometric parameters and aerodynamic parameters.

The {take-off variables} category includes the  air density at the take-off location and the runway length. The first can be measured locally using a portable meteorological station, obtained from the nearest surface meteorological station or, as a last resource,  from the local climatology provided by the national weather service.

The {geometric parameters} category includes the wing area, the wingspan, the height of the wing above the ground, the aircraft empty weight and the weight of the cargo. The rolling friction between the tires and the runway must also be  known. It is modeled according to $F=\mu_R$ \cite{atrito}, where $\mu_R$ is the rolling friction coefficient obtained from the literature (\cite{anderson} and \cite{roskam}, for example) or experimentally, as shown in \cite{atrito_experimento} and \cite{pellegrini_alves}. Another important input variable is the relation between the propulsion system thrust  and the velocity of the airplane, i.e., the thrust curve. 
In general, it is modeled as a second degree polynomial, whose coefficients  can be obtained by curve fitting to wind tunnel  data or using the low cost method proposed in \cite{pellegrini_rodrigues}. 

Finally, in the aerodynamic parameters category, the drag polar curve is the most important input information. The curve is composed by values of the lift coefficient and their associated drag coefficient values, for every possible angle of attack, $\alpha$.  The polar curve is obtained by methods widely disseminated in the aeronautical area, such as \cite{anderson} and \cite{roskam}. The category also includes the drag and lift coefficients for stall angle and take-off angle, and the Oswald efficiency factor.

The calculations for each phase are described below. In each case, only the main output variables are mentioned. The complete list can be found in Appendix A.

\subsection{Take-off}
Taking off an aircraft consists of accelerating it from rest to a speed where the lift is greater than its weight. The required runway length is evaluated in this phase. The lift force is obtained through the well-known relationship:
\begin{equation} \label{eq:lift}
F_L=\frac{1}{2}\rho C_L V^2 A
\end{equation}
where $\rho$ is the air density, $C_ {L}$ is the lift coefficient (which depends on the angle of attack of the wing), $V$ is the speed of the aircraft in relation to the air and $A$ is the projected wing area. 

The acceleration is created by the thrust force of the engine, $T$, which must be greater than the sum of the two dissipative forces that act on the aircraft during take-off, namely the total drag:
\begin{equation} \label{eq:drag}
F_D=\frac{1}{2}\rho C_{D} V^2 A
\end{equation}
 and the rolling resistance, 
\begin{equation} \label{eq:drag}
F_R=\mu_r(W- F_L) 
\end{equation}
where $C_ {D}$ is the drag coefficient for the entire aircraft, $N$ is the normal force, and $\mu_R$ is the combined rolling resistance coefficient of the complete  landing gear.

According to the classic aeronautical literature (\cite{anderson}, for example), the dependence of the engine thrust on the aircraft's velocity relative to the air, for the kind of propulsion system used, can be modeled by:
\begin{equation} \label{eq:tracao}
T=\alpha V^2 + \beta V + T_0
\end{equation}
where $\alpha,$ $\beta$ and $T_0$ are  constants specific for each propulsion system, $T_0$ being known as the static thrust.

For the analysis of take-off, it is necessary to  calculate  the take-off velocity, $V_ {to}$, equaling the total weight of the aircraft, $W$, to the lift force under takeoff conditions and inserting the usual 1.2 safety factor \cite{far}. Therefore:
\begin{equation}  \label{eq:Vlo}
V_{to}=1.2\sqrt{\frac{2(W/A)}{ \rho C_{Lstall}}}
\end{equation}
where $(W / A)$ is the wing load, $C_{Lstall}$ is the maximum lift coefficient, occurring at the associated $\alpha_{stall}$ angle of attack. It is to be attained by the airplane near the end of the runaway when the pilot \emph{rotates} the airplane, pitching its nose up. Using this value, the runway length required for take-off, $S_{to}$, may be obtained  from the closed analytical model proposed by \cite{pellegrini_rodrigues}:
\begin{multline} \label{eq:decola1}
S_{to}  =  \frac{W }{ 2gB} \Bigg[ \ln \left| \frac{BV_{to}^2 + CV_{to} + D }{ D} \right| \\
- \frac{2C }{ \sqrt{\Delta_1}}\bigg(\! \arctan\frac{2BV_{to}+C}{\sqrt{\Delta_1}} - \arctan\frac{C }{ \sqrt{\Delta_1}} \bigg) \Bigg]
\end{multline}
for ${\Delta_1=4BD-C^2}>0$, and 
\bigskip\bigskip\bigskip
\begin{multline} \label{eq:decola2}
S_{to}  =  \frac{W }{ 2gB} \Bigg[ \ln \left| \frac{BV_{{to}}^2 + CV_{to} + D }{ D} \right| \\
 - \frac{C }{ \sqrt{\Delta_2}}
 \bigg(\!\ln\frac{2BV_{to}+C - \!\sqrt{\Delta_2}}{2BV_{to}+C + \!\sqrt{\Delta_2}} 
 \cdot\frac{C + \!\sqrt{\Delta_2}}{C - \!\sqrt{\Delta_2}}  \bigg)\Bigg]
\end{multline}
for ${\Delta_2 = C^2-4BD}>0$, where 
\begin{equation} \label{eq:A}
B = \alpha - \frac{\rho}{2} (C_{D} - \mu_r C_L) S 
\end{equation}
\begin{equation} \label{eq:B}
C = \beta  - \mu_r C_L S 
\end{equation}
\begin{equation} \label{eq:C}
D = T_0 -\mu_rW 
\end{equation}

The last result is not found in textbooks. It was recently proposed in substitution of the classic averaged approach, presented for example in \cite{anderson}, where the necessary integration is simplified using average values of the forces involved, calculated at 70\% of $V_{to}$. Such a simplification yields good practical results but obscures the understanding of the take-off process and is limited by several factors \cite{perkins}. The most relevant is that the approximated integration does not yields a $S=S(V)$ relation. Nevertheless, the value of $S_{to}$ according to \cite{anderson} is also calculated by APT as a reference.

\subsection{Climb}

The climb phase consists of raising the aircraft's altitude. The process takes place after the take-off so that the aircraft can reach the desired altitude. In this analysis, the maximum climb rate, ${C\!R}_ {max}$, and the maximum climb angle, $\theta_ {max}$, are calculated using the following equations: 
\begin{equation} \label{eq:RC}
C\!R_ {max}=\left(\frac{T-F_D }{ W}\right)V
\end{equation}
\begin{equation} \label{eq:phiRC}
\theta_{max}=\arcsin\left(\frac{T-F_D }{ W}\right)
\end{equation}

The toolbox also plots the curve ${C\!R}={C\!R}(V)$. The value of ${C\!R}_ {max}$ is obtained through the plot  and does not occurs at $\theta_{max}$.

\subsection{Steady, level flight}
In this phase the aircraft is in a situation of horizontal flight, at a constant speed. Thus, the lift force must equal the weight, and the thrust generated by the propulsion system must equal  the drag. The maximum speed that the aircraft can reach, the cruising speed, the maximum range and maximum range speed can  be calculated by equating the values of available and required power,  $P_A$ and $P_R$, i.e, by putting 
\begin{equation} \label{eq:power}
P_A = {TV}
\end{equation}
\begin{equation} \label{eq:power}
P_R = {F_D V}
\end{equation}
The resulting dependencies $P_A = P_A (V)$ and $P_R = P_R (V)$ are presented graphically to the user to allow a better analysis of the aircraft's behavior under different conditions.

\subsection{Level turn}
It is a simple curve maneuver, in which the aircraft performs a rolling movement around its longitudinal axis, tilting its wings in relation to the ground, while moving in a horizontal plane. The vertical component of the lift is equal to the weight of the aircraft and the horizontal component creates the centripetal acceleration necessary to travel along a circular path. For this  situation the maximum angle of roll and the minimum radius of curvature are calculated, respectively, by
\begin{equation} \label{eq: phimax} 
\phi_ {max} = \cos ^ {- 1} \left (\frac {2(W / A)} {\rho\, C_ {Lstall} V_ {max} ^ 2}  \right) 
\end{equation} 
and
\begin{equation} \label{eq: Rmin} 
R_ {min} = \frac {2(W / A)} {\rho g C_ {Lstall}}
\end{equation} 
where  $C_ {Lmax}$ is the maximum lift coefficient and  $V_ {max}$ is maximum aircraft speed.

\subsection{Gliding}
It is a situation of descent with the propulsion system in idle, that is, with zero thrust. It is used in the landing phase of small aircraft or in case of engine failure. The angle and the rate of descent are calculated for maximum autonomy and maximum reach, respectively,
\begin{equation}
\theta_ {min} = \tan ^ {- 1} \left (\frac {1} {(C_L / C_D) _ {max}} \right) 
\end{equation}
where $(C_L / C_D) _ {max}$ is the maximum aerodynamic efficiency.

\subsection{Landing}
It consists in decelerating the aircraft from the speed with which it touched the runway, to the rest. In general, thrust is considered null. Other forces related to the use of brakes, lift spoilers or the reverse\footnote{A configuration in which the propulsion system generates negative thrust} to shorten the  length necessary for landing, are not considered here and, thus, the result is conservative. Here, APT uses the same formulation employed for the take-off with $T = 0$, yielding  
\begin{equation} \label{eq: landing} 
S_l = \frac{W }{ 2gB} \Bigg [\ln \left (\frac{BV_l ^ 2 + D }{ D} \right) \Bigg] 
\end{equation} 
where 
\begin{equation} \label{eq: Vl}
V_ {l} = 1,3 \sqrt \frac{2(W / A) }{ \rho C_{Lstall}} 
\end{equation} 
according to (\cite{far}), and 
\begin{equation} \label{eq: landing}
 B = {\frac {\rho} {2} (C_ {D} - \mu C_ {L}) A} 
\end{equation} 
\begin{equation} \label{eq: land} 
D = {\mu W} 
\end{equation}

\subsection{Payload variation}
The payload is the weight that the aircraft is capable of carrying at different altitudes at which take-off can occur, since the variation in altitude leads to a variation in air density, consequently altering the thrust. 
the lift and the drag. This load is evaluated through an iterative process, in which the take-off distance is repeatedly calculated for increasing masses, until it exceeds the runway length (with a prescribed factor of safety).  The whole process is then repeated for various altitudes, with their respective values of  density. The result, given by the toolbox in graphic form, shoes the dependence of the payload with altitude. 

\subsection{Flight envelope}
The flight envelope is the region of the speed vs. altitude graph in which the aircraft can sustain straight and level flight. As altitude increases, stall, maneuver and minimum speeds increase, however, maximum speed decreases. The envelope traced by taking  these speeds into account, delimits the aircraft's operating range. In this analysis, this envelope is presented graphically.

\section{Software Design}
The Matlab\textsuperscript{\textregistered}  platform was the chosen development environment due to its renowned ability to deal with  large numbers of variables and to perform calculations with large indexed variables.  The free student version was choosen.

Matlab\textsuperscript{\textregistered} is  an interactive environment specialized in numerical and algebraic manipulation that uses as its basic variables  matrices that do not require dimensioning. The language is rather similar to standard mathematical writing, which makes coding  easier than in  languages as  FORTRAN, C and Java. Matlab\textsuperscript{\textregistered} possesses a tool to aid developing graphic interfaces called GUIDE (Graphical User Interface Design Environment) that enables the construction of the layout and automatically generates the associated code, that can be further edited to add other desired functionalities. 

The numerical data output can be easily written in Matlab\textsuperscript{\textregistered} by the omission of the semicolon at the end of an equation or using the command \emph{disp}. The graphic data output can be done via the \emph{plot} command, which has a very intuitive syntax. Even after one graph has been made, the user still has wide possibilities to change its details, altering titles, legends, axes, fitting curves, colors, etc. 

Finally, in spite of the fact that Matlab\textsuperscript{\textregistered} is a commercial software, due to its huge numerical  problem solving capability  and simplicity of coding, it is widely spread in the academic community, being available at most universities in its student edition and higher. To ensure that the didactical  objectives of the tool were fulfilled, the user interface was designed in a simple and intuitive way. Thus, students with little knowledge of airplane performance and even of aeronautics are able to carry out the analyses quickly and correctly.  It is only necessary that the student enter the required data  and select "perform analysis", and all graphs and values are calculated and presented. Also, a large number of explanatory remarks where included in the source code, for those with access to it.  For users with only access to the executable, the program was externally documented in an organized manner. 

For the development of the {toolbox}, several {scripts} were initially implemented to carry out the necessary performance analyzes. Each {script} contains the equations necessary to perform one of the analyzes, as well as the reading commands for the input data and writing or plotting commands for the output data. The set of {scripts} works interactively to allow the output data from one analysis to be used as input data from another.

\begin{figure*}[ht]
	\centering
	\includegraphics[width=0.9\textwidth]{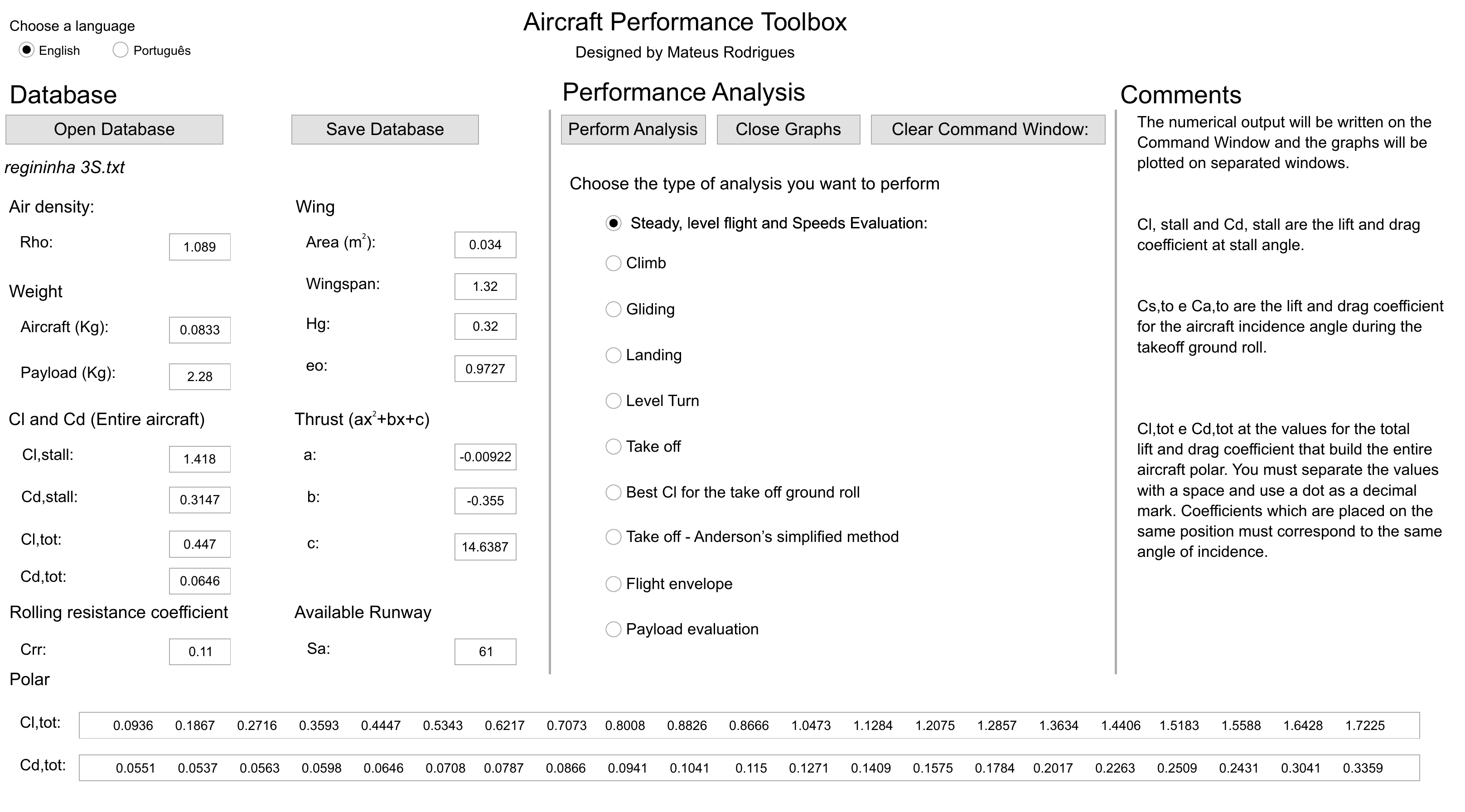}
	\caption{Graphic interface.}
	\label{fig:interface}
\end{figure*}

A graphical interface was also created to allow the user to create a database designed to feed the {scripts} and to make their execution easy and intuitive, as shown in Figure \ref{fig:interface}. New databases can be created simply by filling in the blanks with the value of the variables. A previously created bank can be loaded via the `Load Database' button. After inserting a data set, the user has the option of saving them to a text file using the `Save Database'' button.

Once the database is available, the user must select the type of analysis he wants to perform by clicking on the corresponding option. By clicking on the button `Perform analysis' the data entered in the interface at that moment will become available to the system.

Before the code executes the {script} corresponding to the analysis, it will verify that all fields have been filled in, and if not, a  message will appear in the field for observations. The code will also check whether the indexed variables, drag coefficient, and total aircraft lift, have the same number of elements, a crucial detail for the calculations. If not, the user will also be alerted in the manner previously mentioned.

After the execution of the {script} the numeric output data is presented in the Matlab command window and the graphs are displayed each in its window. The user can save the graphic in one of several supported formats: png, jpg, bmp, pdf, etc.

At this point, the user can request to perform other types of analysis, make changes to the database or perform the same analysis again. He can also load another database or simply clear the output data using the `Close graphs' and `Clear Command Window' buttons. 

If the user chooses to perform more than one analysis without closing the existing graphs, the graphs that relate to the same quantities will be displayed in the same window. This feature will allow the user to graphically compare two different aircraft or to analyze the variation of the output parameters given the change in an input parameter. This will be exemplified in the case study presented in the following section.

\begin{figure*}[htb]
	\centering
	\includegraphics[width=0.9\textwidth]{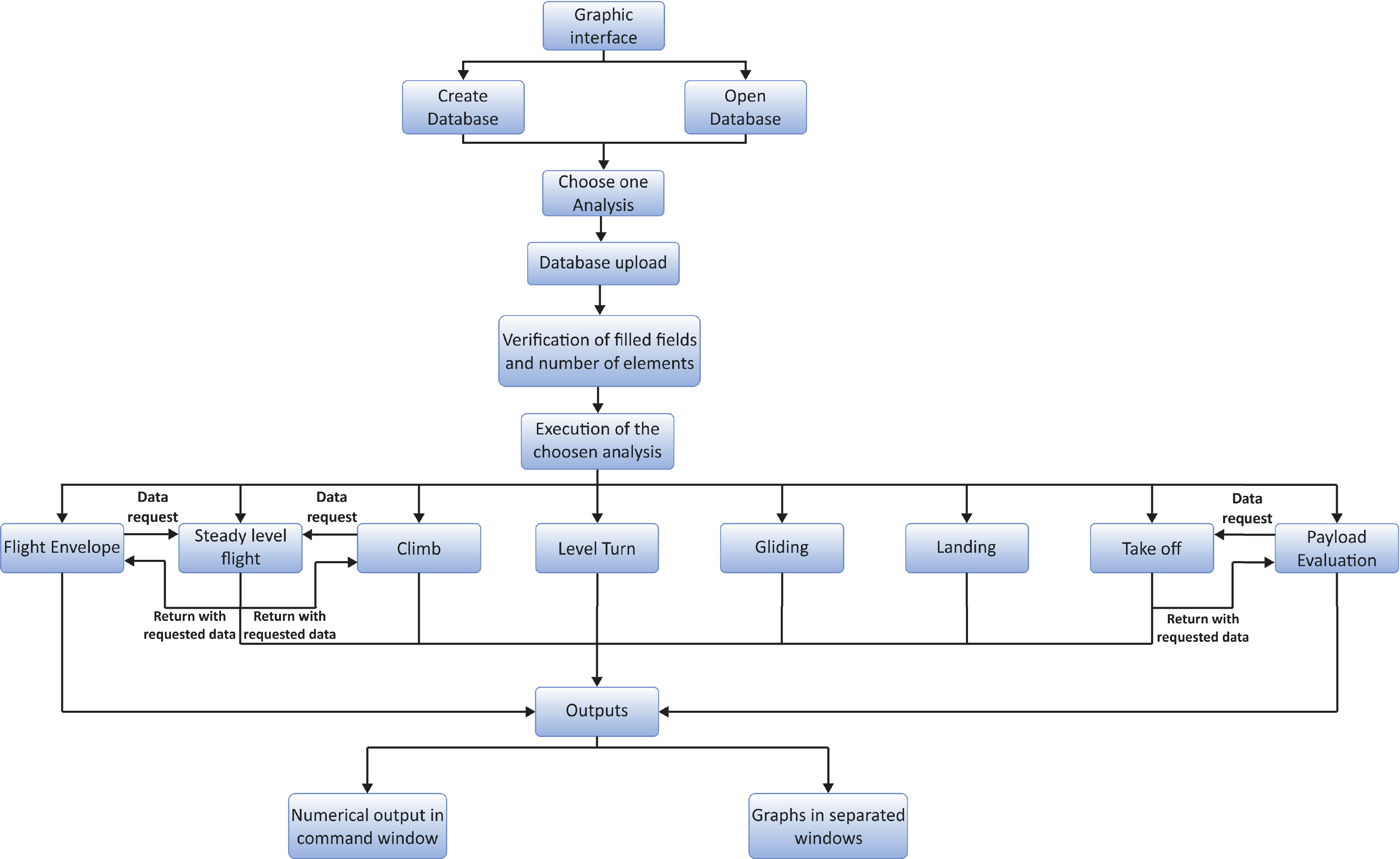}
	\caption{Block diagram.}
    	\label{fig:diagram}
\end{figure*}

For better visualization of the  structure of the APT, a block diagram was built, which can be seen in Figure \ref{fig:diagram}. 

\section{A case study}
APT allows the user to quickly assess the influence of changes in design data on the performance of an aircraft by making changes to the databases. This means that it can be used not only for the evaluation of an existing aircraft but also during the conceptual design phase. 

\begin{figure}
	\centering
	\includegraphics[width=0.40 \textwidth]{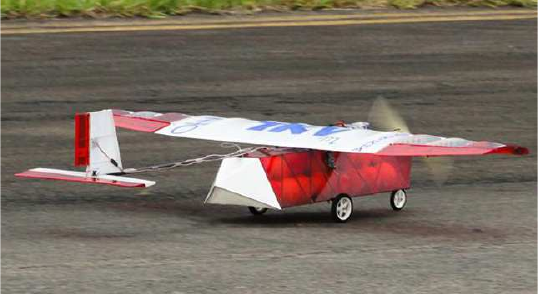}
	\caption{UAV used at 16th SAE Brazil Aero Design competition by the Trem Ki Voa team.}
	\label{fig:aero}
\end{figure}

To illustrate this type of use, this section will show the process of selecting a battery fot the UAV designed  by the Trem Ki Voa Micro team (Figure \ref{fig:aero}) for the 16th SAE Brasil Aero design Competition. Initially, a brief description of the competition challenge will be presented and in the end, the most significant graphics and results will be shown.

The UAV propulsion system consists of a brushless electric engine driven by a frequency inverter, commercially called ESC ({Electronic Speed Control}), and powered by a Lithium-Iron battery. Because  of the weight restrictions imposed by the competition regulations, the least possible weight battery that allows the necessary flight performance must be chosen.

The thrust generated by the kind of engine in question increases as larger electrical voltage is applied to its terminals. For this, the voltage supplied by the battery must be increased. This is possible by increasing the number of cells in it, as they are connected in series and have, by default, 3.3 V each. However, the greater the number of cells, the greater their weight.

The APT was  used to compare the performance of the UAV built by the team, using two 1300 mAh batteries, one with 3 cells in series (called the 3S configuration) and the other with 4 cells in series (the 4S configuration), each weighing, respectively, 148 and 195 g. For comparison, two databases were used, differing only in the thrust coefficients and the empty weight of the aircraft. The values were 0.833 kgf for the 3S battery and 0.880 kgf for the 4S.

The coefficients for the propulsion system thrust curve,  given in Eq. \ref{eq:tracao}, were obtained by the low cost methodology as
\begin{equation}
\text{Configuration  3S:}
\begin{cases}
\alpha=-0.00922 		\text{ N/(m/s)}^2 \\
\beta=-0.355 		\text{ N/(m/s)} \\
T_0=14.387		\text{ N} 			
\end{cases}
\end{equation}
\begin{equation}
\text{Configuration  4S:}
\begin{cases}
\alpha=-0.0110 		\text{ N/(m/s)}^2 \\
\beta=-0.427 		\text{ N/(m/s)} \\
T_0=17.904 		\text{ N}			
\end{cases}
\end{equation}

The other input variables, equal for both configurations, are shown in Table \ref{Tab:input_variables}, where $M_P$ represents the mass of the payload, $C_ {L}^{stall}$ and $C_ {D}^{stall}$ are  the lift and drag coefficient at stall angle, $C_ {L}^{to}$ and $C_ {D}^{to}$ are the lift and drag coefficients at take-off angle, $b$ is the wingspan, $H_{g}$ is the height of the wing in relation to the ground,  $e_o$ is Oswvald's efficiency factor and $S_ {run}$ is available runway length.
\begin{table}[htb]
\small\sf\centering
\caption{Input variables.\label{Tab:input_variables}}
\begin{tabular}{lcc}
\toprule
Variable				&Unit		 				&Value	\\
\midrule
$\rho$   				&kg/m\textsuperscript{3}			& 1.089	\\
$M_P$                 			&kg						& 2.280	\\
$C_ {L}^{stall}$			&-- 						& 1.418	\\
$C_ {D}^{stall}$       		&--						&  0.314	\\
$C_ {L}^{to}$          		&--						& 0.447	\\
$C_ {D}^{to}$         		&--						& 0.064	\\
$\mu_R$                  		&--						& 0.110	\\
$A$           				&m\textsuperscript{2} 			& 0.340	\\
$b$ 	                   			&m 						& 1.320	\\
$H_{g}$              			&m 						& 0.320	\\
$e_o$                       		&--						& 0.972	\\
$S_{run}$                 		&m 						& 61.0		\\
\bottomrule
\end{tabular}\\[16pt]
\end{table}

\begin{figure}[htb]
	\begin{center}
		\includegraphics[width=0.47 \textwidth]{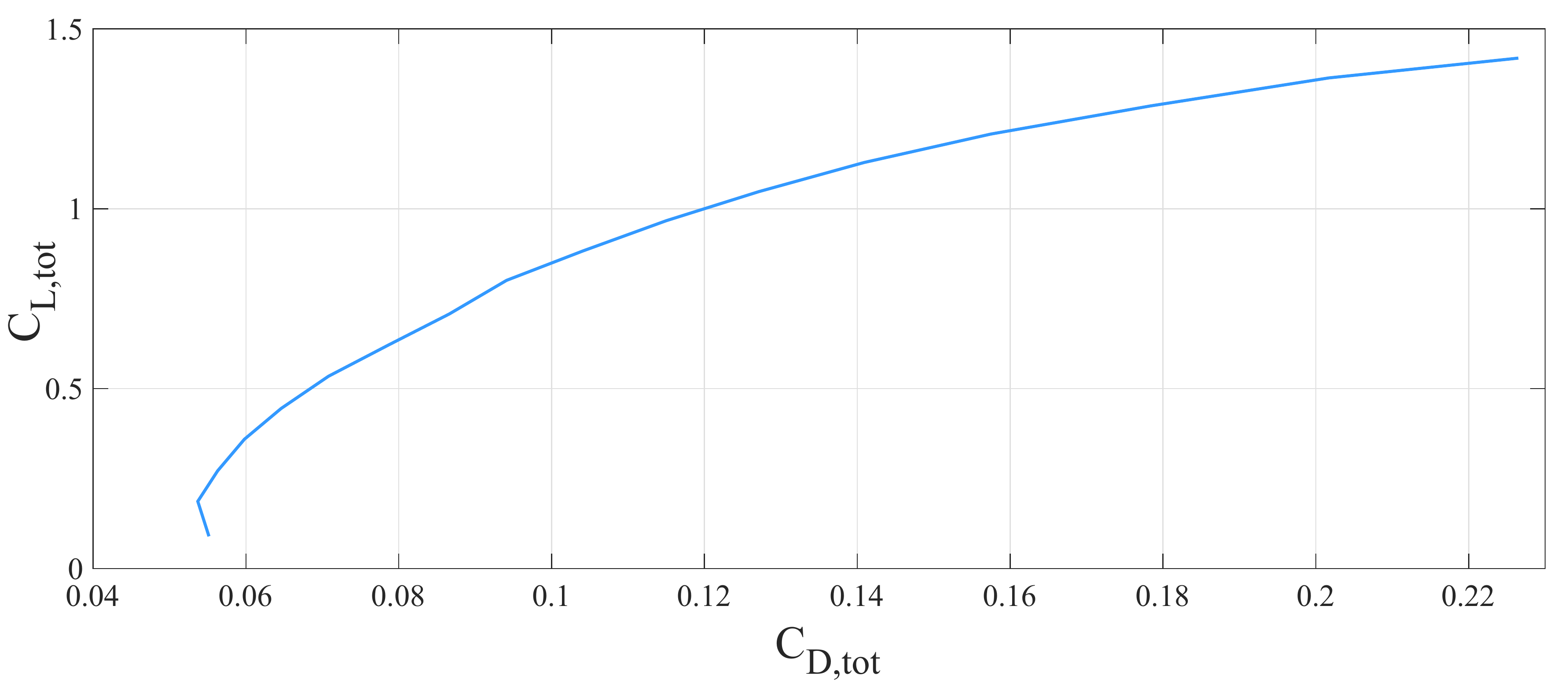}
		\caption{drag polar curve.}
		\label{Fig:polar}
	\end{center}
\end{figure}

The polar curve used in the calculations is shown in Figure~\ref{Fig:polar}.

Some output data for the tested configurations are shown below. Figure \ref{fig:nivelado}, from the analysis of straight and level flight, shows the dependence of  available and required thrust with velocity for both configurations. It is evident that the available thrust shows considerably higher values for the 4S configuration, but required thrust values are only slightly larger. The increase in available thrust is due to the increase in the battery's output voltage and presents a considerable gain in the change from the 3S to the 4S configuration. The required thrust, however, does not show such a strong dependence on the weight of the set, as shown in the figure. Thus, it is clear that, from the aircraft's performance point of view, the 4S configuration is advantageous, increasing the surplus power for maneuvers and the maximum flight speed, as seen in the table \ref{Tab:output_variables}.
\begin{figure}[htb]
	\begin{center}
	\includegraphics[width=0.47 \textwidth]{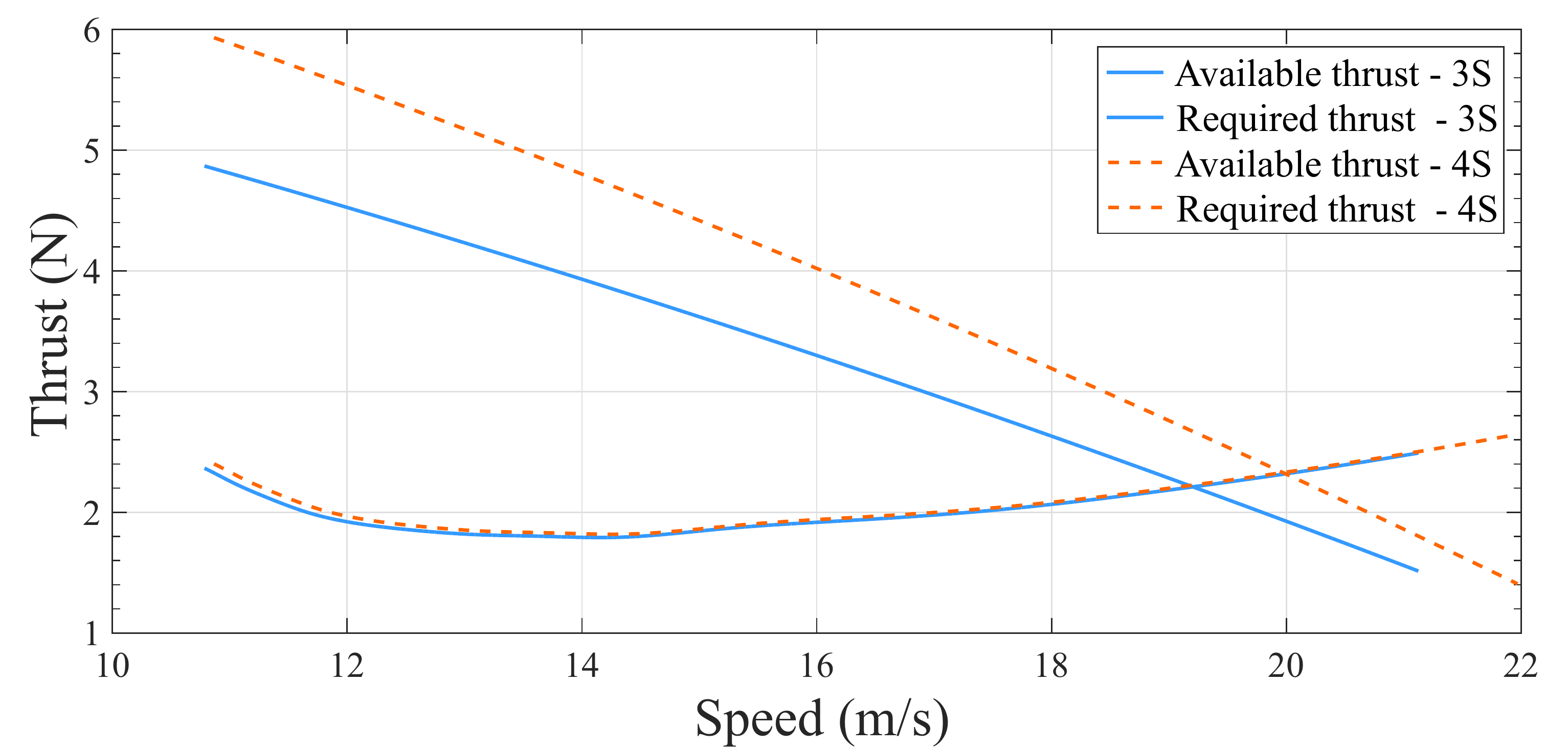}
	\caption{Available and required thrust.}
	\label{fig:nivelado}
	\end{center}
\end{figure}

\begin{figure}[htb]
	\begin{center}
	\includegraphics[width=0.47 \textwidth]{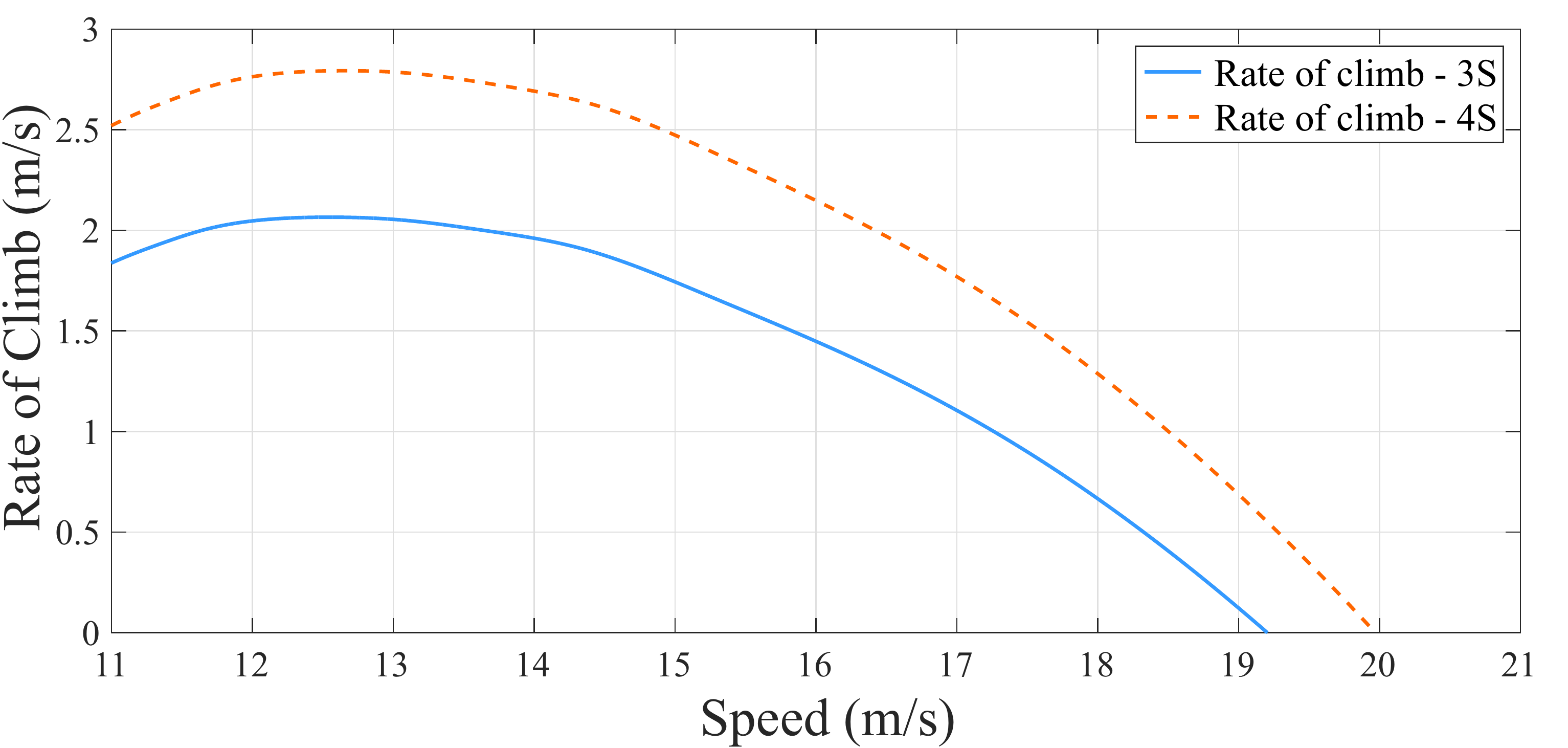}
	\caption{Rate of climb.}
	\label{fig:subida}
	\end{center}
\end{figure}

Figure \ref{fig:subida} comes from the climb analysis and shows the curves of the aircraft's maximum climb rate. Once again, it is noticed that despite the slight increase in weight, the 4S configuration gives the aircraft  a higher climbing capacity, due to the increase in thrust surplus.

\begin{figure}[htb]
	\begin{center}
	\includegraphics[width=0.47 \textwidth]{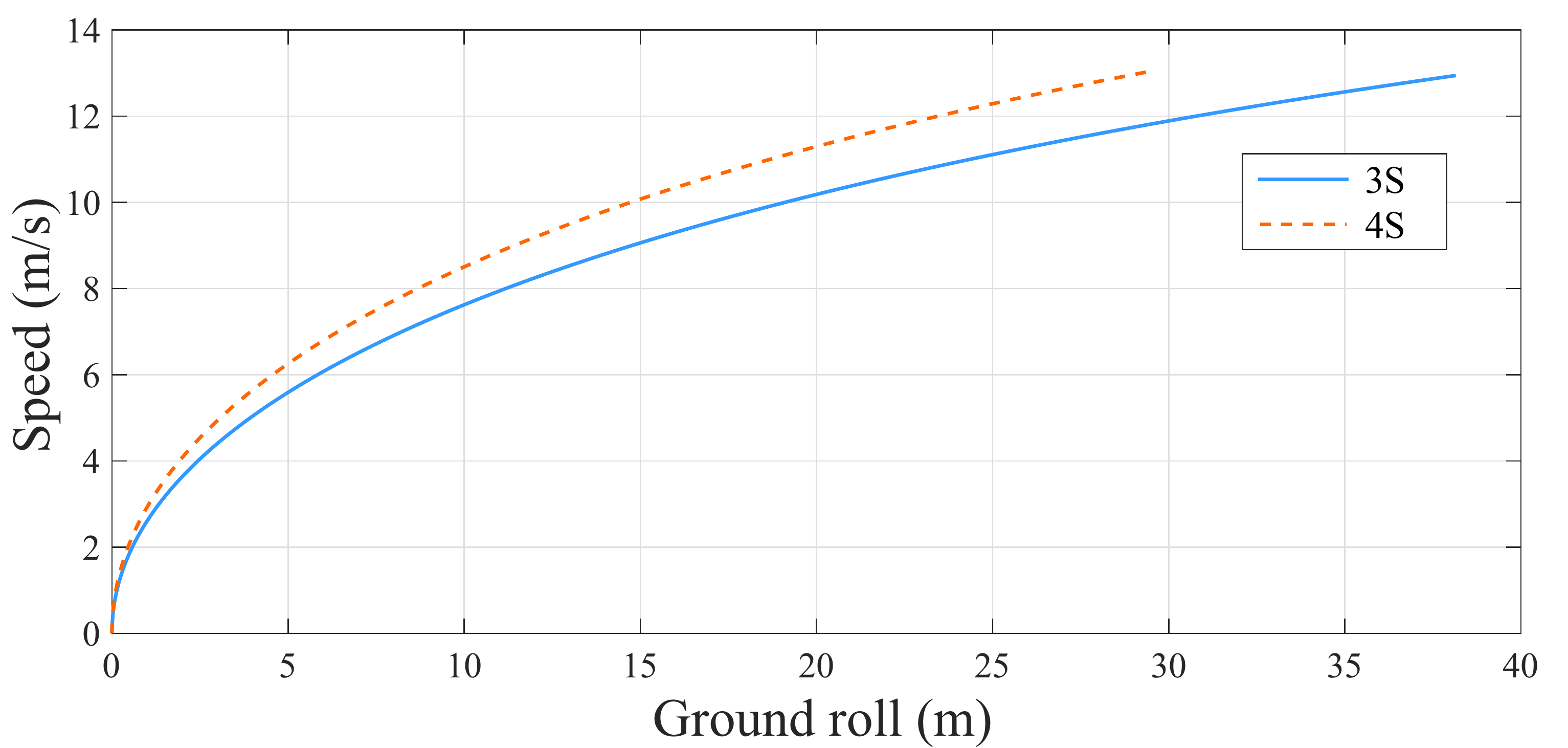}
	\caption{Aircraft speed along the take-off ground roll.}
	\label{fig:decolagem}
	\end{center}
\end{figure}

Figure \ref{fig:decolagem} comes from the take-off analysis and shows the speed along the runway, from rest to take-off speed. Although the 4S configuration demands a slightly larger take-off speed, the aircraft can reach this speed in a distance 23\% smaller than with the 3S configuration.

\begin{figure}[htb]
	\begin{center}
	\includegraphics[width=0.47 \textwidth]{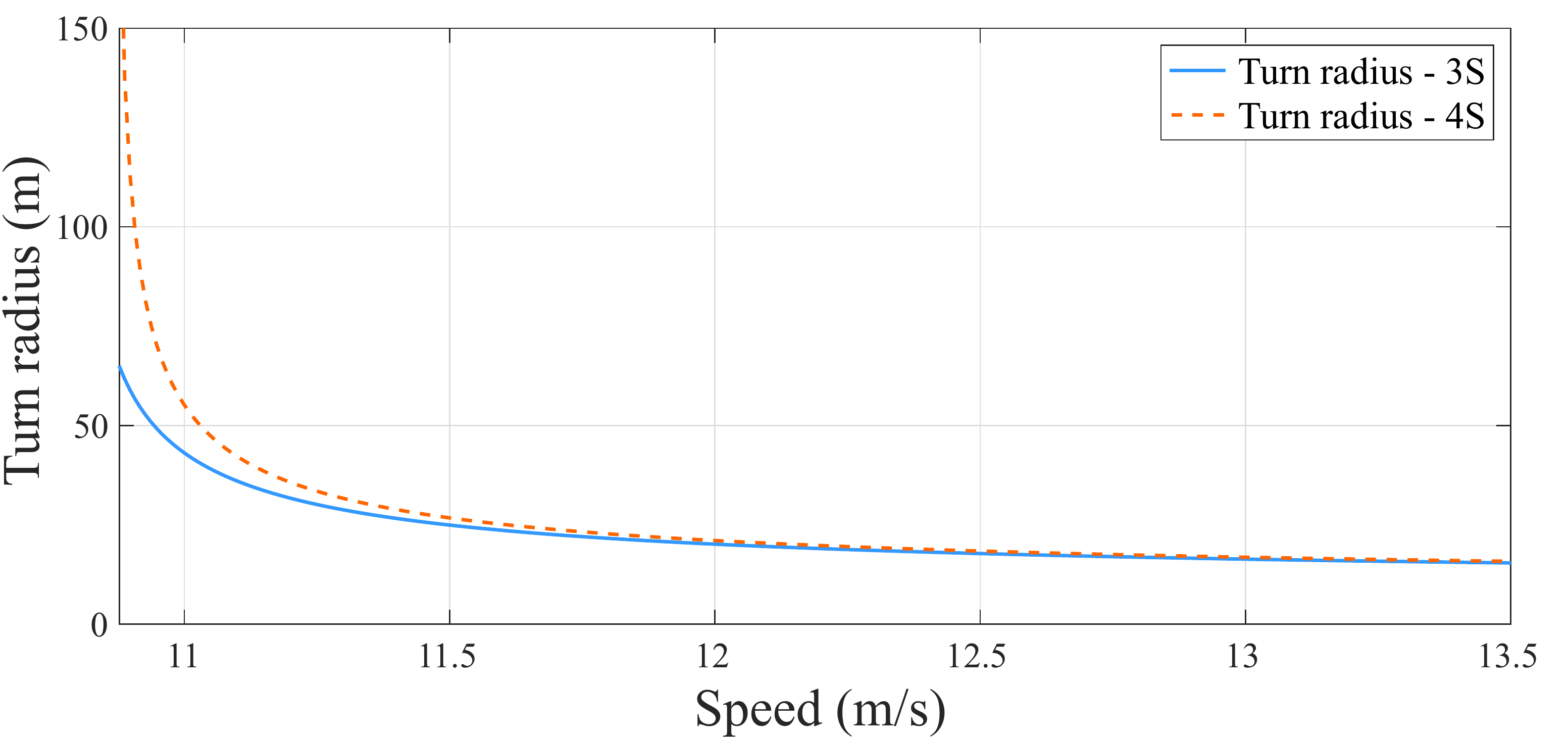}
	\caption{Minimum turning radius.}
	\label{fig:curva}
	\end{center}
\end{figure}

Figure \ref{fig:curva} shows the dependence between the minimum radius of the level curve and the speed for the two configurations. It is clear that the lower power configuration allows for more tight turns for a given speed, due to its lower weight. This is a very important fact for the aircraft's maneuverability and one  the only negative points in the 4S configuration. However, the 4S configuration allows for an overall smaller minimum radius, as it allows the aircraft to reach a higher maximum speed.

\begin{figure}[htb]
	\begin{center}
	\includegraphics[width=0.47 \textwidth]{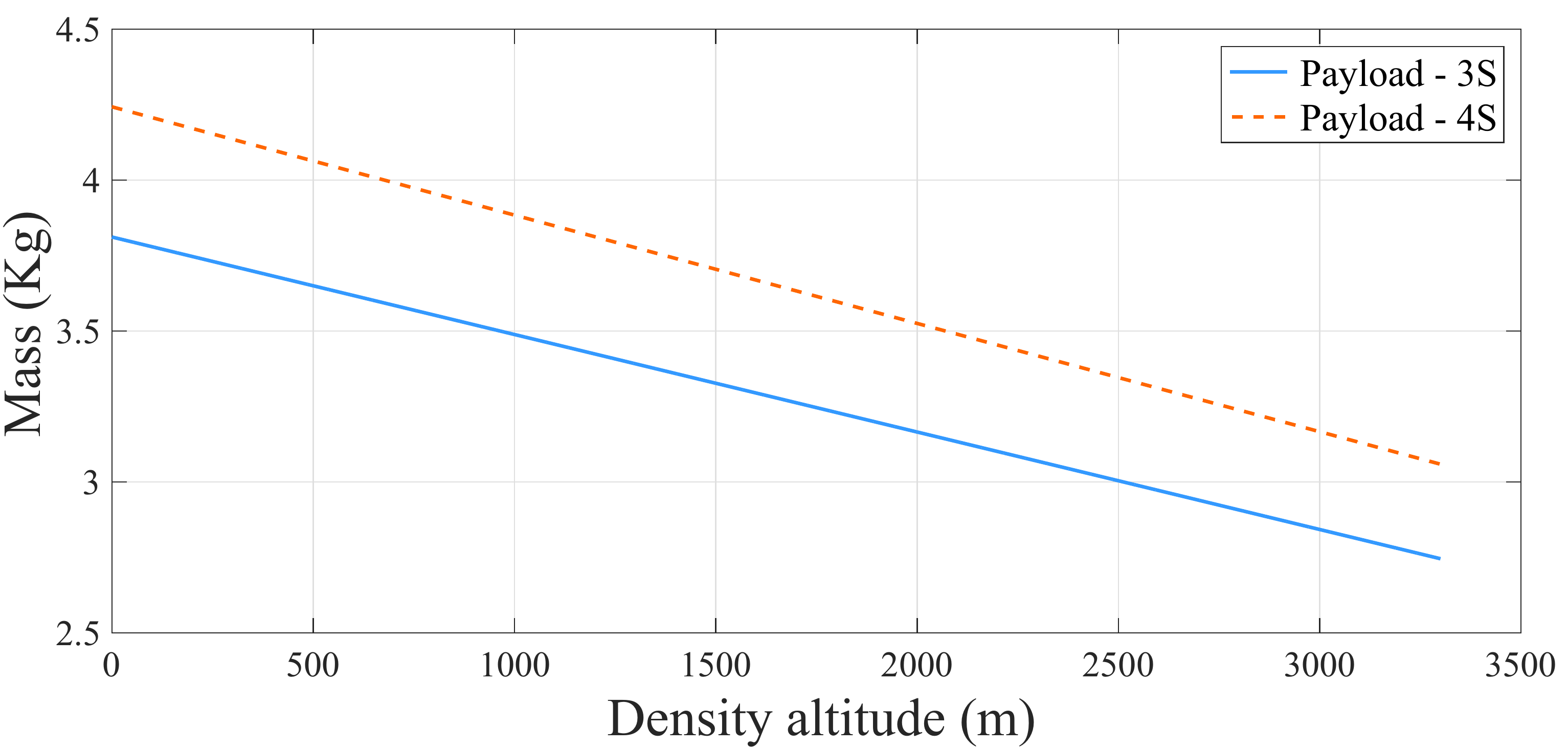}
	\caption{Payload prediction chart.}
	\label{fig:util}
	\end{center}
\end{figure}

Finally, Figure \ref{fig:util} shows the load capacity for both configurations. It is evident that, due to the increase in thrust, the aircraft powered by the 4S battery has a considerably larger load capacity than the aircraft with the 3S battery, an increase of  11\%.

Table \ref{Tab:output_variables} shows the main aircraft output data for various analyses. Here, $V_{stall}$ is the stall speed, $V_{max}$ is   the maximum speed for steady, level flight, $V_{cru}$ is the cruise speed, $\varphi_{max}$ is the maximum climb angle and $\phi_{max}^{rng}$ is the  maximum range angle during gliding, $R_{Dmax}^{rng}$ is the descent reason for maximum range during gliding, $n_{Cmax}$ is the maximum load factor in curve, $R_ {min}$ is the minimum radius in curve,  $\theta_ {max}$ is the maximum roll angle, $S_{to}$  is the runway length needed to take-off and  $S_l$ is  and the runway length needed to land. 
\begin{table}[htb]
\small\sf\centering
\caption{Main output variables.\label{Tab:output_variables}}
\begin{tabular}{lccc}
\toprule
Output variables			&Unit			&3S 			&4S 		\\
\midrule
{$V_{stall}$} 			&m/s			&10.78		&10.86	\\
{$V_{max}$} 			&m/s			&19.20		&19.96	\\
{$V_{cru}$} 				&m/s			&17.28		&17.97	\\
{$R_{Cmax}$} 			&m/s  			&2.06			&2.79		\\
{$\varphi_{max}$} 			&$({}^0)$ 		&9.89			&13.43	\\
{$\phi_{max}^{rng}$} 		&$({}^0)$ 		&6.68			&6.68		\\
{$R_{Dmax}^{rng}$}		&m/s     		&-1.63		&-1.65	\\
{$n_{Cmax}$}			&-			&1.30			&1.48		\\
{$R_{min}$}				&m			&18.37		&16.29	\\
{$\theta_{max}$}			&$({}^0)$ 		&40.14		&47.55	\\
{$S_{to}$}				&m			&38.10		&29.44	\\
{$S_{l}$}				&m			&84.23		&85.51	\\
\bottomrule
\end{tabular}\\[16pt]
\end{table}

It can be seen that almost all the performance parameters indicate the 4S configuration to be advantageous, except for the aforementioned minimum radius of the level curve and for the parameters resulting from glide and landing analysis, where the engine thrust is set to zero. As the 3S configuration would receive, according to the competition rules, only a few more points due to its lower weight, the team opted to use the 4S battery. The decision proved to be correct and the team won the 2014 competition. 

In subsequent years, the team also won the 2016, 2017 and 2019 competitions  using the APT toolbox, and is now four times Brazilian champion. It was also awarded a second overall place in the SAE East AeroDesign world competition of 2015 and another second overall place in the SAE Brazil AeroDesign  competition of 2020. It is now, the most awarded team in Brazil, with four championships, two second places (2012 and 2020) and a world second place.

\section{Didactic Experience}
To access the usefulness of the tollbox, a survey was conducted among students who work and have worked in the performance area of the team, to learn their experiences in the analysis before and after using the {software}. Unfortunately, the success of the tool meant that just a few students had experience before and after its creation. Indeed, from 2015 onward, no student  ever made the calculations by hand again. 

The research was carried out through the Google forms platform. Google forms allow users to "collect and organize information large and small for free". The responses to a survey are stored in spreadsheets (Google Sheets) and can be viewed in graphs or even raw in the spreadsheet. There are different styles of questions and input methods for answers, as well as section breaks, the possibility of uploading files, displaying images or videos, and other features. 

According to the 8 forms received, students think that the {toolbox} considerably speeds up the laborious calculations and helps organize the output data, making it easy to test different aircraft configurations in the initial phase of the project. In addition, it is believed to help in the preparation of the final report and presentations. The tool was also credited with providing greater interaction between the areas, as it helps to organize the input and output data.
 
Another interesting feature mentioned by the students is that the toolbox can be used both in the forward and  backward directions. In the forward direction, it is used for its original purpose, calculating performance characteristics of the UAV from the input data. In the reverse direction, it may be used to calculate input data from measured performance characteristics. 

An example of this, is the possibility of calculating the value of the rolling resistance coefficient for a new set of tires, from the take-off distance measured experimentally . 

The only  disadvantage of APT detected in our diagnosis was that, as students understand that the {code} is able making  the calculations ``on its own'', they tend not to study the theory in depth as they should. This have lead to two  problems in the past years:  (i) students do not acquire the proper understanding about the subject; (ii) students spend a long time using the toolbox in trail and error tests, just to arrive at  conclusions that were obvious  from the theory (that they did not studied properly!). 

Unfortunately, this situation seems to be commonplace in modern engineering courses, whenever a software is available to help in the calculations, and is rather worrying. In a  research  conducted more than one decade ago, \cite{mina} describes characteristics and study habits of first-year students of Iowa State University’s electrical engineering course. Their findings, very consistent with our own observations, identifies 15 common traits and behaviors of the 21\textsuperscript{st} century students. One of then was particularly associated with the problems just mentioned:
\begin{quote}
Students do not seek to find understanding, only answers. As a result, the students are resentful of the confusion that arises naturally from, and is necessary for, the learning process to be successful – that is, to foster understanding, and create lifelong learners\ldots While the answer is important, the understanding of its discovery, or meaning,
is of little value to them --- that one has obtained an answer is sufficient.
\end{quote}

Another observation closely related  with what we consider a misuse of the computational tools is:
\begin{quote}
Students lack an understanding of the learning process. In effect, the students are attempting to gain theoretical
knowledge\ldots merely by working through numerous examples. Indeed, many students seem to believe that the process of learning consists only of working through such examples. 
\end{quote}

And last but not least:
\begin{quote}
Students lack an understanding of the meaning of hard work. Generally, unless they are captivated by a subject, they will not undertake it in earnest. Consequently, they are unwilling to perform the routine\ldots necessary to develop their skills.
\end{quote}

As for future versions of the toolbox, students showed interest in a module to calculate manual launch take-offs. The development of toolboxes for other areas of the project, with the capacity of interaction  between them, was also mentioned and is already underway. Priority was requested in the future development of a module for a complete aerodynamic analysis toolbox, capable of generating the polar drag curve, to remove the need for the user to enter the  data manually. An important, although obvious, detail about this request is that it depends on the availability of dedicated members of the Team, with adequate training in MatLab\textsuperscript{\textregistered} and other necessary software. Of course, being a member is not mandatory, but in our experience is highly recommended, as it generally guarantee the necessary theoretical basis.  Indeed, APT was developed by one of the authors, a second year Electrical Engineering undergraduate student, member of the performance sector.

\section{Conclusions}

We believe that APT was able to meet all the requirements stipulated initially, including the didactic ones. The code was implemented in Matlab\textsuperscript{\textregistered},  available in the vast majority of Brazilian universities. The interface created allows the user to create and modify input databases that  can be used to evaluate the performance characteristics of different aircraft designs, in an easy and intuitive way. As shown in the case study, the data output is also made in a simple and clear way, allowing for a quick comparison between the different  configurations studied. This is a much important  quality, once time is restricted in the kind of competition APT was designed to help with. 

The evaluations of the tool returned by its users were considered to be extremely positive. The only inconvenience, i.e., the tendency of the students to treat the software as a black box, is being currently addressed but it seems to be part of a much larger problem: the 21\textsuperscript{st} century students study habits. This is an issue that will certainly  have to be dealt with, as computational  tools like the APT each day occupy a larger share  of the  available tools for complex or laborious calculations. Our recommendation for colleagues  who are thinking of creating and implementing similar tools, is to consider this question in as an early stage of the development of the tool as possible.  

Regarding the results obtained with the aid of  APT, the  titles  conquered by the Team in the  competitions may be considered a good indicator of its success. Of course, the group relied on many other skills to obtain those titles, but the software was certainly one important factor in the process, as pointed out by most students involved.

\section*{Acknowledgments} 

\addcontentsline{toc}{section}{Acknowledgments} 

The authors would like to thank the Department of Thermal and Fluid Sciences and the Departments of Electrical and Mechanical Engineering at the Federal University of São João da Rei for the infrastructure provided. They also thank the other members of the Trem Ki Voa Micro Team for their help and support in the 2014---2020 c ompetitions.

This work was partially supported by FAPEMIG (Minas Gerais State  Agency for Research and Development)  under grant TEC-APQ-03108-13. 

\section*{Appendix A - Output data by type of analysis}

The following tables show all the output variables, listed by type and, in the case of graphical variables, the parameter according to which they are plotted (Depend.). Symbols that have not been defined throughout the text will be defined before the table where they will be inserted.

Table \ref{Tab:output_decolagem} presents the output data of the take-off analysis, in which the symbols have the following meaning. 
$V_ {inst}$ is the instantaneous aircraft speed during take-off; $T_ {p}$ is the track position; $V_v$ main wind speed; $W_p$ weight of the transported cargo.

\begin{table}[htb]
\small\sf\centering
\caption{Take-off.\label{Tab:output_decolagem}}
\begin{tabular}{lcccc}
\toprule
Output variables		&Unit		&Type			&Dependency 	&Unit		\\
\midrule
{$S_{to}$} 			&m 		&Numerical		&- 			&-		\\
{$V_{inst}$} 		&m/s		&Graphical		&$T_{p}$ 		&m		\\
{$S_{to}$} 			&m		&Graphical		&$V_v$ 		&m/s		\\
{$W_{p}$} 			&N 		&Graphical		&$S_{to}$ 		&m		\\
\bottomrule
\end{tabular}\\[10pt]
\end{table}

Tables \ref{Tab:output_steady} and \ref{Tab:output_climb} show the output data of the analysis of straight and level flight and climb, respectively.  In which the symbols have the following meaning: $V_{div}$ diving speed, $V_{man}$ maneuver speed, $V_{aut}^{max}$ maximum autonomy speed, $V_{rch}^{max}$ maximum reach speed, $T_{d}$ thrust available, $T_{r}$ required thrust, $T_{lo}$ leftover thrust and $P_{A}$ available power, $P_{R}$ power required, $\theta_{climb}$ climb angle.

\begin{table}[htb]
\small\sf\centering
\caption{Steady, level flight.\label{Tab:output_steady}}
\begin{tabular}{lcc}
\toprule
Output variables 		&Type 		&Dependency 	\\
\midrule
{$V_{stall}$} 		&Numerical 		&- 			\\
{$V_{to}$} 			&Numerical 		&- 			\\
{$V_{max}$} 		&Numerical 		&- 			\\
{$V_{cru}$} 			&Numerical 		&- 			\\
{$V_{div}$} 			&Numerical 		&- 			\\
{$V_{man}$} 		&Numerical 		&- 			\\
{$V_{aut}^{max}$} 	&Numerical 		&- 			\\
{$V_{rch}^{max}$} 	&Numerical 		&- 			\\
{$T_{d}$} 			&Graphical 		&Vel. 			\\
{$T_{r}$} 			&Graphical 		&Vel. 			\\
{$T_{to}$} 			&Graphical 		&Vel. 			\\
\bottomrule
\end{tabular}\\[16pt]
\end{table}

\begin{table}[htb]
\small\sf\centering
\caption{Climb.\label{Tab:output_climb}}
\begin{tabular}{lcc}
\toprule
Output variables 				&Type 		&Dependency 	\\
\midrule
{$R_{Cmax}$} 				&Numerical		&- 			\\
{$(R/C)_{max}$}				&Numerical 		&- 			\\
{$\varphi_{max}$}				&Numerical 		&- 			\\
{$\varphi_{max}$}				&Numerical 		&- 			\\
{$P_{A}$}					&Graphical 		&Vel.			\\
{$P_{R}$}					&Graphical 		&Vel.			\\
{$\theta_{climb}$} 				&Graphical 		&Vel.			\\
\bottomrule
\end{tabular}\\[10pt]
\end{table}

Table \ref{Tab:output_turn} shows the output data from the curve analysis. The symbols have the following meaning: $n$ load factor; $R$ curve radius; $\theta$ scroll angle.

\begin{table}[htb]
\small\sf\centering
\caption{Level turn.\label{Tab:output_turn}}
\begin{tabular}{lcc}
\toprule
Output variables 		&Type 		&Dependency 	\\
\midrule
{$n_{Cmax}$} 		&Numerical 		&- 			\\
{$R_{min}$}			&Numerical		&- 			\\
{$\theta_{max}$}		&Numerical		&- 			\\
{$n$} 				&Graphical		&Vel. 			\\
{$R$} 			&Graphical 		&Vel. 			\\
{$\theta$} 			&Graphical 		&Vel. 			\\
\bottomrule
\end{tabular}\\[10pt]
\end{table}

Table \ref{Tab:output_gliding} shows the output data from the glide analysis, where the symbols have the following meaning. $\phi_ {max, aut}$ angle of maximum autonomy in glide; $RD_ {max, aut}$ descent rate for maximum gliding autonomy, glide speed: $V_{gli}$, $RD$ descent rate; $\phi$ gliding angle. 
\begin{table}[htb]
\small\sf\centering
\caption{Gliding.\label{Tab:output_gliding}}
\begin{tabular}{lcc}
\toprule
Output variables								&Type			&Dependency	\\
\midrule
$\phi_{max}^{aut}$							&Numerical		&- 			\\
$RD_{max}^{aut}$							&Numerical 		&- 			\\
$V_{gli}$ \footnotesize\textrm{for}$RD_{max}^{aut}$		&Numerical		&- 			\\
{$\phi_{max}^{rch}$}							&Numerical		&- 			\\
{$RD_{max}^{rch}$}							&Numerical		&- 			\\
{$V_{gli}$ \footnotesize\textrm{for} $RD_{max}^{alc}$}		&Numerical		&- 			\\
{$RD$}									&Numerical		&Vel. 			\\
{$\phi$}									&Graphical		&Vel 			\\
\bottomrule
\end{tabular}\\[16pt]
\end{table}

Tables \ref{Tab:output_landing}, \ref{Tab:output_payload} and \ref{Tab:output_envelope} show the output data for the landing analysis, payload evaluation and flight envelope, respectively.

\begin{table}[htb]
\small\sf\centering
\caption{Landing.\label{Tab:output_landing}}
\begin{tabular}{lcc}
\toprule
Output variables 		&Type 		&Dependency 	\\
\midrule
{$S_{l}$}			&Numerical		&- 			\\
\bottomrule
\end{tabular}\\[16pt]
\end{table}

\begin{table}[htb]
\small\sf\centering
\caption{Payload variation.\label{Tab:output_payload}}
\begin{tabular}{lcc}
\toprule
Output variables		&Type			&Dependency 	\\
\midrule
{Payload prediction}	&Graphical		&Altitude 		\\
\bottomrule
\end{tabular}\\[16pt]
\end{table}

\begin{table}[htb]
\small\sf\centering
\caption{Flight envelope.\label{Tab:output_envelope}}
\begin{tabular}{lcc}
\toprule
Output variables 		&Type			&Dependency	\\
\midrule
{Flight envelope}		&Graphical		&Altitude		\\
\bottomrule
\end{tabular}\\[16pt]
\end{table}
\vfill



\end{document}